\documentclass[reprint, amsmath, amssymb, aps, showkeys, nolongbibliography, numerical]{revtex4-2}

\usepackage[usenames,dvipsnames]{xcolor}
\usepackage[mathlines]{lineno}

\usepackage{float}
\usepackage{graphicx}
\usepackage[colorlinks=true, linkcolor=blue, allcolors=blue]{hyperref}
\usepackage[mathlines]{lineno}
\usepackage{bm}
\makeatletter
\newsavebox{\@brx}
\newcommand{\llangle}[1][]{\savebox{\@brx}{\(\m@th{#1\langle}\)}%
  \mathopen{\copy\@brx\kern-0.5\wd\@brx\usebox{\@brx}}}
\newcommand{\rrangle}[1][]{\savebox{\@brx}{\(\m@th{#1\rangle}\)}%
  \mathclose{\copy\@brx\kern-0.5\wd\@brx\usebox{\@brx}}}
\makeatother

\begin{document}
\title{Vortex lattice melting and critical temperature shift in rotating Bose-Einstein condensates}

\author{Julián Amette Estrada$^{1,2}$}\email{julianamette@df.uba.ar}
\author{Marc E.~Brachet$^{3}$}\email{brachet@phys.ens.fr}
\author{Pablo D. Mininni$^{1,2}$}\email{mininni@df.uba.ar}
\affiliation{$^1$Universidad de Buenos Aires, Facultad de Ciencias Exactas y Naturales, Departamento de Física, Ciudad Universitaria, 1428 Buenos Aires, Argentina,}
\affiliation{$^2$CONICET - Universidad de Buenos Aires, Instituto de F\'{\i}sica Interdisciplinaria y Aplicada (INFINA), Ciudad Universitaria, 1428 Buenos Aires, Argentina,}
\affiliation{$^3$Laboratoire de Physique de l'{E}cole {N}ormale {S}up\'erieure, ENS, Universit\'e PSL, CNRS, Sorbonne Universit\'e, Universit\'e de Paris, F-75005 Paris, France.}

\date{\today}

\begin{abstract}
We investigate a shift in the critical temperature of rotating Bose-Einstein condensates mediated by the melting of the vortex lattice. Numerical simulations reveal that this temperature exhibits contrasting behavior depending on the system configuration: a negative shift occurs for fixed trap potentials due to the expansion of the condensate, while a positive shift is observed for fixed volumes, where vortex lattice rigidity suppresses thermal fluctuations. We introduce a vortex-energy model that captures the role of vortex interactions, the positional energy of the vortex lattice, as well as the phase transition and how the vortex lattice disappears. The findings provide insights into the thermodynamic properties of rotating condensates and the dynamics of vortex lattice melting, offering potential parallels with other quantum systems such as type-II superconductors.
\end{abstract}


\maketitle

\section{Introduction}

One of the hallmarks of quantum fluids is the existence of quantized vortices, first theorized by Onsager \cite{Onsager1949} and Feynman \cite{Feynman1955}. Since then, quantum vortices have been extensively studied. Recently, advances in experiments and simulations have sparked growing interest in rotating Bose-Einstein condensates (BECs), as they provide a platform to link quantum gases and fluids with type-II superconductors and other quantum materials \cite{makinen2023}. In these systems, the presence of an external order field causes vortices to form an array known as the Abrikosov lattice, which affects the order parameter. For an infinite gas, Tkachenko demonstrated that the lattice must be triangular to minimize free energy \cite{Tkachenko1965}. Such lattices and their onset have been observed experimentally \cite{AboShaeer2001,Madison2001,Schweikhard_2004}. Once the lattice forms, the previously three-dimensional state of the system becomes quasi-two-dimensional, and the system's behavior near equilibrium is dominated by vortex dynamics, with waves acting as lattice perturbations \cite{Coddington2003}.

The effect of vortex lattices in a BEC critical temperature remains largely unexplored. It is evident that increasing the temperature must imply the disappearance of the lattice, as eventually there must be no condensate phase remaining. However, the melting of vortex arrays has been studied in detail. As quasi-long range order develops in the vortex crystal, the theory for phase transitions in 2D systems developed by Kosterlitz and Thouless, as well as Halperin, Nelson and Young, is applicable \cite{Kosterlitz1973,Halperin1978}. Gifford and Baym \cite{Gifford2008} studied the dislocation-mediated thermal melting of a vortex lattice in a rotating superfluid using elasticity theory. The process was described meticulously, and the melting temperature was obtained for the homogeneous case. 

Experimentally, the melting of two-dimensional (2D) vortex lattices in superconductors was studied in \cite{Guillamn2009}, revealing hexatic and smectic-like phases. Additionally, the melting of a vortex array through dislocations in a quasi-2D BEC experiment was recently examined in \cite{Sharma2024}. Numerically, Monte Carlo simulations have also been employed to investigate vortex behavior in three-dimensional (3D) systems in \cite{Hetzel1992, Kragset2006}. In \cite{Hetzel1992} a frustrated 3D XY model was considered, revealing a first-order phase transition for the melting of unpinned Abrikosov lattices in type-II superconductors. In \cite{Kragset2006} vortices under cylindrical confinement were studied, showing that fluctuations concentrate near the condensate borders and that the vortex lattice melts from the outside in. 

In rotating BECs, melting of the vortex lattice always occurs close to or below the Bose-Einstein critical temperature \cite{Gifford2008}. As a result, if the vortex lattice can still be observed at a given temperature, it provides a lower bound to the critical temperature. This can be qualitatively understood as follows: In a rotating BEC the lattice appears as the result of the impossibility of the superfluid phase to rotate as a rigid body. The lattice requires the spatial long range order of the condensate, and its presence evidences the condensate existence. Moreover, for rotation frequencies sufficiently below the trap frequency in elongated traps, the melting temperature is close to the condensate critical temperature (see details in \cite{Gifford2008}, and recent results in \cite{Sharma2024}). Finally, in that regime melting of the vortex lattice is induced by thermal fluctuations instead of by quantum fluctuations, which is also the case under typical experimental conditions \cite{Gifford2008}.

In this article we study how the critical temperature of a cylindrically trapped rotating BEC changes under different conditions, and the role of the vortex lattice in this process. We consider the regimes more relevant for most experiments, with small enough rotation frequencies and in systems dominated by thermal fluctuations.

\section{Rotating Bose-Einstein condensates at finite temperature}

A few methods are available study the equilibrium and dynamics of interacting Bose gases at finite temperature, including regimes up to the critical temperature to study phase transitions. These methods can be separated into classical field methods (as, e.g., the projected or truncated stochastic Gross-Pitaevskii or Ginzburg-Landau equations, depending on whether the dynamics or just equilibria are sought for \cite{Berloff2002, Blakie2005, Davis2006, Berloff2007, Shukla2019}), and full stochastic Gross-Pitaevskii of quantum Boltzmann formulations that can properly describe quantum fluctuations \cite{Gardiner_2002, Calzetta_2007} (see \cite{Proukakis2008} for a detailed comparison). The former methods assume that the lowest energy modes are sufficiently populated to be described classically, while the latter provide a quantum field description (although, in practice, most of their numerical implementations are also limited to classical distributions \cite{Proukakis2008}). Here we use the Ginzburg-Landau equation to obtain states at zero temperature, and a truncated stochastic Ginzburg-Landau formulation to generate classical field states at finite temperature, in all cases considering the effect of rotation.

The Hamiltonian that describes the order parameter $\psi$ of a rotating BEC at zero temperature is
\begin{eqnarray}
    \mathcal{H} = \int d^3r &\Bigg[& \frac{\hbar^2}{2m} |\boldsymbol{\nabla} \psi|^2 + \frac{g}{2} |\psi|^4 + \nonumber \\
    {} &&
    V(\bm{r}) |\psi|^2 - \psi^* (\bm{\Omega} \cdot \bf{J}) \psi \Bigg],    
    \label{eq: hamiltonian}
\end{eqnarray}
where $m$ is the bosons mass, $g$ is proportional to the s-wave scattering length, $V(\bm{r})$ is the external potential, $\bm{\Omega}=\Omega \hat{z}$ is the rotation angular velocity, and $\bf{J}$ is the angular momentum. Its variation gives the well known Gross-Pitaevskii equation, whose stationary solutions at a given energy can be obtained from the evolution of the Rotating Ginzburg-Landau equation (RGLE),
\begin{eqnarray}
    \frac{\partial \psi}{\partial t} &=& \left[  \frac{\hbar}{2 m}\nabla^2 - \frac{g}{\hbar} |\psi |^2 - \frac{V(\bm{r})}{\hbar} + \frac{\bm{\Omega} \cdot \bf{J}}{\hbar} + \frac{\mu}{\hbar} \right]\psi ,
    \label{eq: RGLE}
\end{eqnarray}
where $\mu$ is the chemical potential. To obtain finite temperature states we follow the same procedure as in \cite{AmetteEstrada2022_AVS}, and we generalize Eq.~\eqref{eq: RGLE} as a Langevin equation,
\begin{eqnarray}
    \frac{\partial \psi}{\partial t} &=& \left[  \frac{\hbar}{2 m}\nabla^2 - \frac{g}{\hbar} |\psi |^2 - \frac{V(\bm{r})}{\hbar} + \frac{\bm{\Omega} \cdot \bf{J}}{\hbar} + \frac{\mu}{\hbar} \right]\psi + \nonumber \\ 
    {} &&
    \sqrt{\frac{2 }{\mathcal{V} \hbar \beta}} \zeta (\bm{r},t) ,
    \label{eq: SRGLE}
\end{eqnarray}
which is the Stochastic Rotating Ginzburg-Landau equation (SRGLE) that provides a classical field model \cite{Berloff2014} in which $\zeta (\bm{r},t)$ is a delta-correlated random process such that $\left< \zeta (\bm{r},t) \zeta^* (\bm{r}',t')\right> = \delta (\bm{r} - \bm{r}') \delta (t-t')$, and $\sqrt{2 / \mathcal{V} \hbar \beta}$ controls the amplitude of fluctuations through a temperature $T \sim 1/\beta$ ($\mathcal{V}$ is the system volume). This equation, when written for a finite number of Fourier modes (i.e., truncated) up to a cut-off wave number $k_\textrm{max}$ using a Galerkin truncation, is equivalent to a Fokker-Planck equation for the state probability $\mathbb{P}[\{\hat{\psi} (\bm{k},t), \hat{\psi}^* (\bm{k},t)\}]$, and converges to thermal states in the grand canonical ensemble. The mass can be fixed instead of $\mu$ (i.e., to obtain canonical ensemble states) by solving an equation for the chemical potential \cite{AmetteEstrada2022_AVS}. In the following we solve those equations for thermal states, and Eq.~\eqref{eq: RGLE} for $T=0$. 

This methodology and similar methods have been used to study the disappearance of Bose-Einstein condensation under many conditions \cite{Berloff2002, Blakie2005, Davis2006, Berloff2007, ClarkDiLeoni2018, Shukla2019, AmetteEstrada2022_AVS}, following the approach described in \cite{Berloff2014} to solve finite temperature dynamics, the route to condensation, and to find critical temperatures. Also, they have been used to study the process of non-equilibrium condensation \cite{Berloff2002, Berloff2007}, and in particular, to determine the shift on critical temperature in condensates comparing successfully with experiments and showing better agreement than other mean field theories \cite{Davis2006}. A review of these methods and of their advantages and disadvantages can be found in \cite{Proukakis2008}.

\begin{figure}
    \includegraphics[width=\columnwidth]{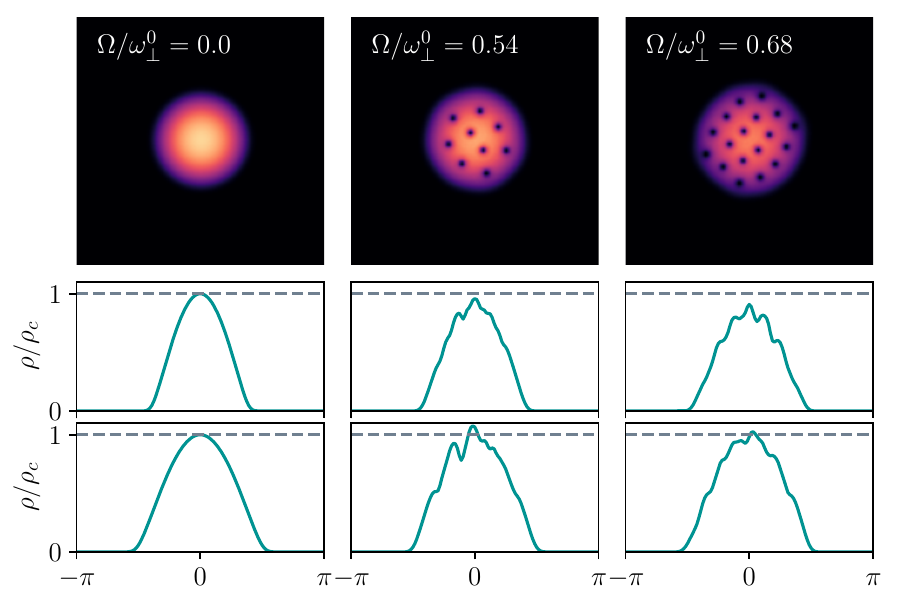}
    \caption{Density in real space for the BEC at zero temperature for different rotation speeds. Top panels show $\rho(z,y,z=0)$. The other rows show $\rho(x, y = 0, z = 0)/\rho_c$ for fixed potential (mid panels) and for fixed volume (bottom panels). }
    \label{Figure:densities}
\end{figure}

\begin{figure}
    \includegraphics[width=\columnwidth]{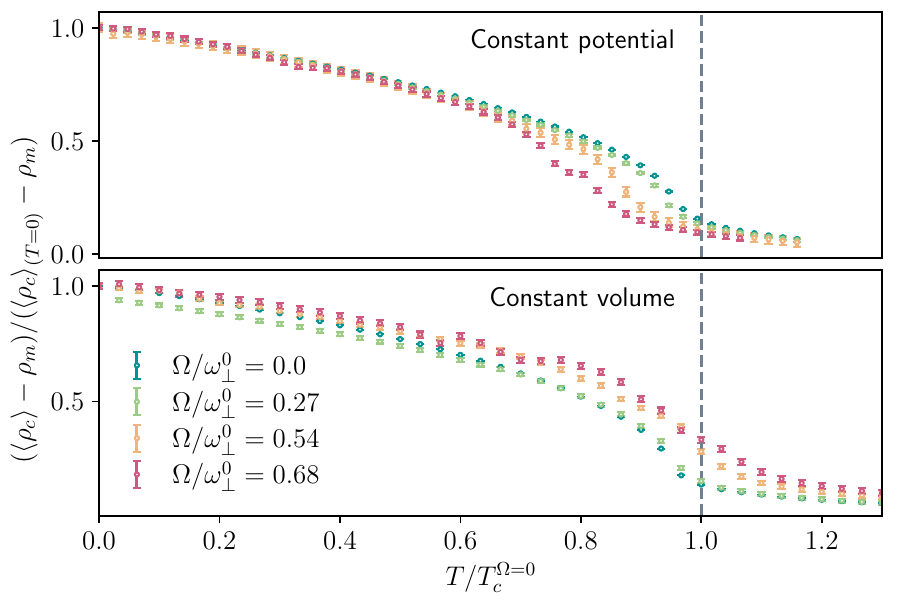}
    \caption{Density at the center of the trap minus the mean density in the trap, normalized by their values at $T=0$, for various rotating speeds (see labels in the insets) as a function of $T/T_c^{\Omega = 0}$. From top to bottom, the cases with fixed potential and fixed condensate volume are shown. The vertical dashed lines indicate the estimated critical temperature in the non-rotating case. Error bars represent $99.7 \%$ confidence intervals of the mean density standard error.}
    \label{Figure:transitions simulations}
\end{figure}

To solve these equations we use an axisymmetric potential $V({\bm r}) = m \omega_\perp^2 (x^2+y^2)/2 $. The system is integrated in a cubic domain of dimensions $[-\pi,\pi]L \times [-\pi,\pi]L \times [-\pi,\pi]L$ with periodicity in $z$, using a Fourier-based pseudo-spectral method with a spatial grid of $N^3 = 128^3$ grid points. The $2/3$ rule is used to control aliasing instabilities, and an implicit first-order Runge-Kutta method is used for time integration with the GHOST parallel code, which is publicly available \cite{Mininni2011}. The non-periodic potential and angular momentum operator are computed using the methods in \cite{AmetteEstrada2022}. Results are presented in units of a characteristic speed $U$, the unit length $L$ (proportional to the condensate mean radius), and a unit total mass $M$. All parameters are fixed by setting the speed of sound as $c = (g \rho_0/m^2)^{1/2} = 2 \, U$, the condensate healing length as $\xi = \hbar/(2\rho_0 g)^{1/2} = 0.00353 \, L$ (these two relations are strictly valid for uniform density condensates, so here they apply locally), the reference trapping frequency to $\omega_\perp^0 = 1.85  \, U/L$, and the unit density as $\rho_0 = 1 \, M/L^3$. Quantities can then be scaled by setting dimensional values for $U$, $L$, and $M$. In experiments, typical values are $L\approx  5 \times 10^{-5}$ m and $c \approx 1 \times 10^{-3}$ m/s \cite{White2014,AboShaeer2001}. This results in $\xi \approx 2.8 \times 10^{-7}$ m and a trap frequency $\omega_\perp \approx 116 $ Hz. Particle densities in experiments with Na atoms are $\approx 10^{14}$ cm$^{-3}$ atoms using $5 \times 10^7$ atoms \cite{AboShaeer2001}, and a total mass to maximum density ratio of $5\times 10^{-13} \, \text{m}^3$; in our simulations this ratio is $1.4 \times 10^{-14} \, \text{m}^3$.

\begin{figure*}
\centering
    \includegraphics[width=\textwidth]{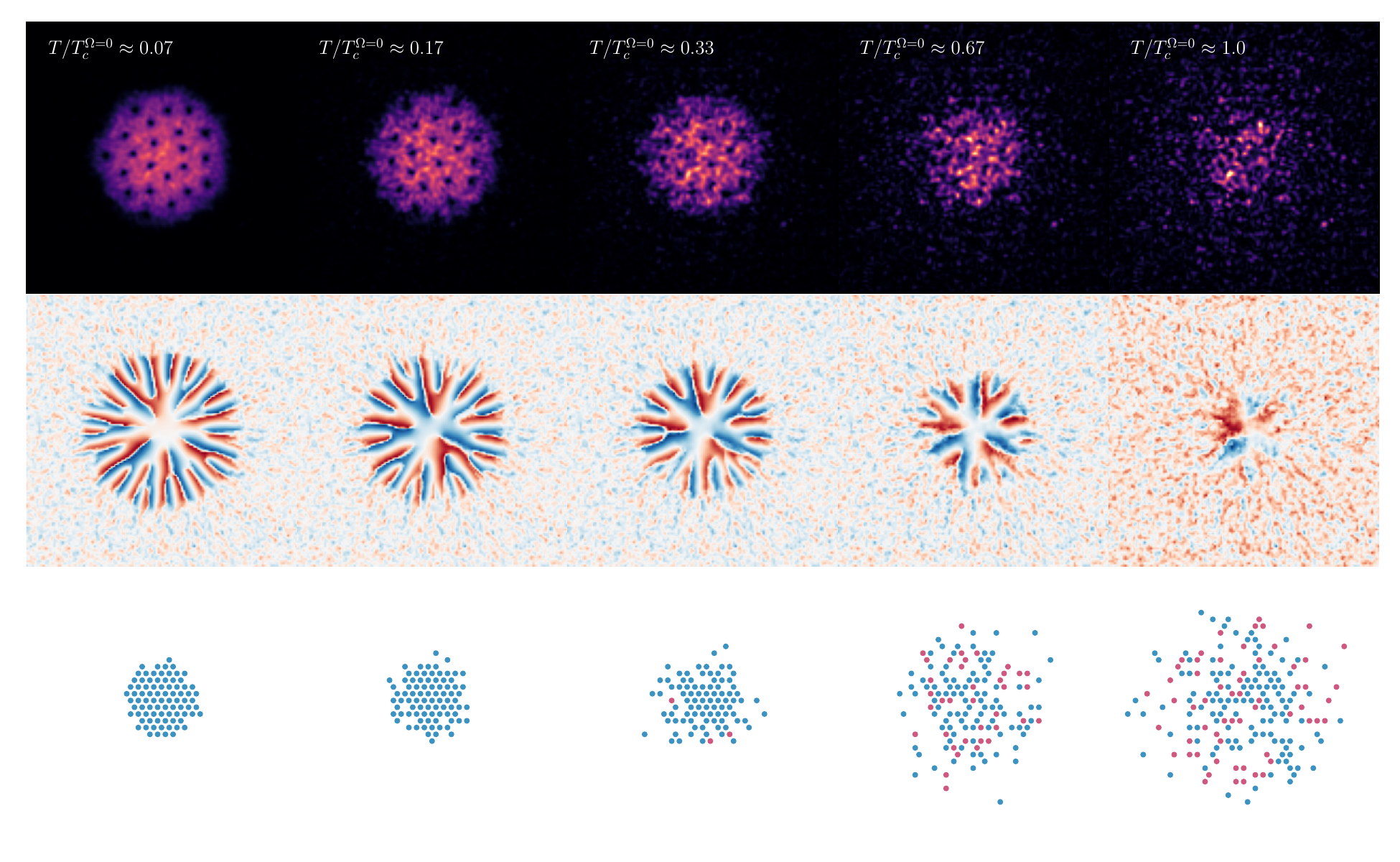}
    \caption{Top: $\rho(x,y,z=0)$ with increasing temperature ($\Omega / \omega_{\perp}^0= 0.68$).  Middle: phase of $\psi$, averaged along the rotation axis, for the same configurations. Blue corresponds to $-\pi$ and red to $\pi$. Phase shifts are caused by quantized vortices. Bottom: Vortex-energy model with increasing $T$. Blue vortices are parallel to the rotation axis, red vortices are anti-paralell.}
    \label{Figure:phase transition}
\end{figure*}

\section{Numerical simulations}

To study the effect of rotation on the critical temperature of the BEC and on lattice melting we must consider two scenarios: the case in which the trap frequency is kept constant, and the case with constant volume. In the former, as $\Omega$ increases the condensate expands (and its volume thus increases) as a result of the centrifugal force. In the latter, the trap frequency must be increased with $\Omega$ in such a way that the Thomas-Fermi radius of the BEC remains fixed, and thus the volume remains unchanged. Note that this behavior results from the fact that rotation produces a change in the effective trap frequency: The third term and part of the fourth term on the r.h.s.~of Eq.~(\ref{eq: hamiltonian}) can be rewritten using an effective confining potential \cite{Fetter_2009} of the form $m (\omega_{\perp}^2 - \Omega^2) r_\perp^2/2$ (with $r_\perp^2 = x^2 + y^2$), and thus we can adjust $\omega_{\perp}$ to keep the effective external potential felt by the condensate the same. Both for constant $\omega_{\perp}$ and for constant volume, the total mass of the condensate is kept constant.

Figure \ref{Figure:densities} shows the density of the BEC in real space at zero temperature, for different rotation rates. The top three panels display the density in the midplane perpendicular to the rotation axis, for different values of $\Omega$ with a fixed trap frequency. As rotation increases, a larger vortex lattice is generated. The two lower rows present $\rho(x, y = 0, z = 0)/\rho_c$, where $\rho_c$ is the density in the center, for constant trap frequency (top) and constant volume (bottom). For constant trap frequency, the central density decreases as the condensate radius grows with increasing $\Omega$. Conversely, with constant volume, the central density remains approximately the same, with variations caused by the vortex lattice.

Figure \ref{Figure:transitions simulations} shows the mean density around the trap center, $\langle \rho_c \rangle$ (the average is taken to account for fluctuations), minus the mean density in the entire trap, $\rho_m$, normalized by their corresponding values at $T=0$, for different $\Omega$ and as a function of the temperature normalized by the critical temperature of the BEC without rotation, $T/T_c^{\Omega = 0}$. Note that the critical temperature $T_c$ is determined as the inflection point of the curves in Fig.~\ref{Figure:transitions simulations}. In the laboratory the local (or optical) central density has been used before to estimate the condensed fraction \cite{Davis1995,Chen2022}. In homogeneous condensates in numerical simulations, the lowest Fourier modes of the momentum can be also used to identify the fraction of particles in the condensate (see, e.g., \cite{Shukla2019}). In the presence of a trap, correlation functions are also used \cite{Blakie2005}, as well as the spectrum of momentum \cite{AmetteEstrada2022}. We verified that these methods yield similar values for $T_c$, and in the following use the local central density to allow for more direct comparisons with experiments (see also \cite{Amette2024_engines} for a discussion).

In all cases in Fig.~\ref{Figure:transitions simulations} the condensate density decreases with $T$ until it reaches the phase transition. The two aforementioned cases are shown in this figure: the case with fixed trap frequency, and the case with fixed condensate radius. Opposites results are obtained: in the former case a negative shift in $T_c$ is seen for increasing $\Omega$, while in the latter case $T_c$ increases as $\Omega$ increases. To understand these differences we must first note that volume, temperature, total mass and rotation speed are the four relevant thermodynamic quantities. In the first case two thermodynamic variables change, while in the second case only one changes. The decrease of $T_c$ with $\Omega$ observed in the first case with fixed potential can then be explained considering the growth of the condensate radius as $\Omega$ increases, leading to a reduction in the central peak density (see Fig.~\ref{Figure:densities}). Lower densities in condensates cause a negative shift in $T_c$. In the second case with constant volume, rotation promotes order and gives the condensate additional resilience to fluctuations, with the vortex lattice seeming to play a crucial role. The two cases share similarities with the behavior observed in trapped condensates with increasing repulsive interaction parameter $g$, where a reduction in $T_c$ results from cloud broadening, while increased interaction at constant density leads to a positive shift in $T_c$ \cite{Smith2011}. 

It could be argued that in the former case, at constant radius (and constant density), the change in $T_c$ can be the result of the change in the trap potential $\omega_\perp$ used to keep the radius constant (even though the effective potential corrected by the centrifugal potential remains the same). We verified that under constant density and for $\Omega=0$, the effect of increasing the potential is actually the opposite to that seen in the bottom panel of Fig.~\ref{Figure:transitions simulations}. It results in a small decrease on $T_c$ (see the Appendix for details). Therefore, the change in the critical temperature must be associated with the rotation. One then may ask: Through which mechanism does rotation affect $T_c$? And what happens to the vortex lattice as $T$ grows?

To answer these questions, we first study the effect of temperature on the lattice. Figure~\ref{Figure:phase transition} shows the condensate for $\Omega / \omega_{\perp}^0= 0.68$ at increasing $T$. The top row shows the mass density in SRGLE simulations, in the midplane perpendicular to the rotation axis. As temperature increases, fluctuations cause vortex positions in the lattice to shift. This is evident by the blurring of the vortices, starting from the borders as $T$ increases (see similar behaviour in \cite{Kragset2006}). Additionally, the condensate shape becomes less defined, making individual vortex identification challenging. The borders of the cloud are the first to deform, whereas the center of the condensate maintains it shape. At large temperatures this behavior can be better appreciated by looking at the phases of $\psi$, which are shown (averaged over the vertical direction) in the middle row of Fig.~\ref{Figure:phase transition}. Vortices correspond to points where the phase around them shifts by $2\pi$ (i.e., the radial origin of blue and red stripes). Note that indeed the border of the lattice melts first as $T$ increases, with the vortices in the center remaining with increasing disorder. 

\begin{figure}
    \includegraphics[width=\columnwidth]{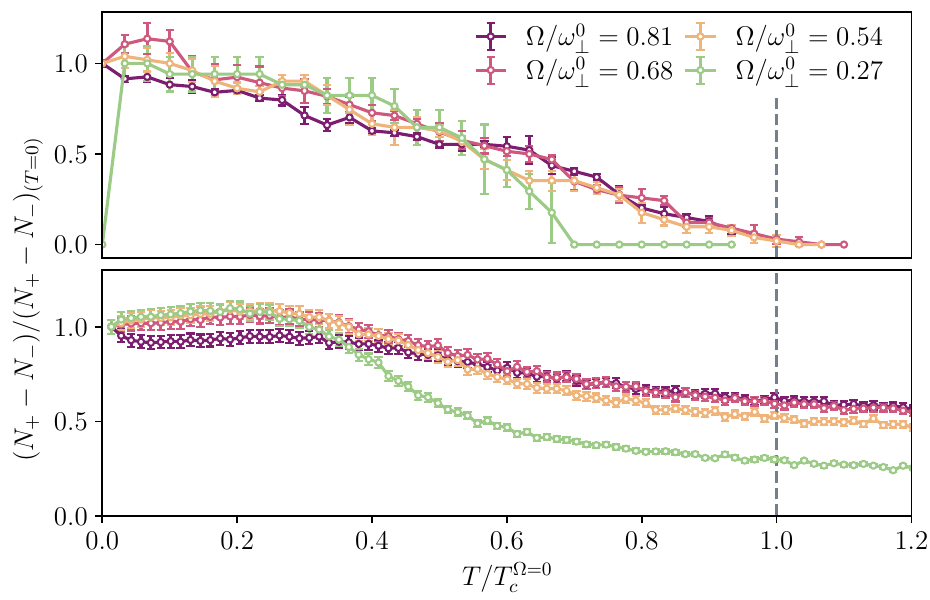}
    \caption{Net number of positive vortices (i.e., aligned with the axis of rotation), normalized by the same number at $T=0$, in RSGLE simulations with constant volume (top) and in the vortex-energy model (bottom), as a function of $T$ and for different $\Omega$. Temperatures in both cases are normalized by the same value of $T_c^{\Omega=0}$. Error bars indicate $95 \%$ confidence intervals of the standard error.}
    \label{Figure:vortex evolution}
\end{figure}

The melting of the lattice as $T$ increases can be further confirmed by studying the number of vortices co-rotating with the condensate as a function of the temperature (see Fig.~\ref{Figure:vortex evolution}, top panel). As $T$ approaches $T_c$, the lattice dissapears. Moreover, an overshoot in the number of vortices is seen at small $T$ in many cases. This is caused by fluctuations induced by temperature, which make new states with a larger number of vortices available with more available energy. Finally, the number of vortices near $T_c^{\Omega=0}$ also increases with $\Omega$. Thus, the lattice persistence, and in particular the need for the condensed phase to maintain this structure in the rotating case, appear to shield the condensate as $T$ increases, effectively raising $T_c$. Fluctuations then concentrate at the edges of the lattice, where the condensate density is lower, and vortices  in that region are the firsts to disappear.

\section{The vortex-energy model}

The vortex array enhances the coherence of the condensate, allowing it to persist at higher temperatures. This leads to the question: can this effect be attributed solely to interactions between vortices within the array? To explore this we construct an Ising-like model for the system, taking into account vortex interactions. In our vortex-energy model we assume an underlying triangular Abrikosov lattice of vortices is present, and define the Hamiltonian of our vortex-energy model as
\begin{eqnarray}
    \mathcal{H}^T = - \frac{1}{2 \pi} \Gamma_0^2 \sum_{\ll ij \gg} \sigma_i \sigma_j ln(r_{ij}) - \alpha h N_c \Omega \sum_i \sigma_i \nonumber
    \\ 
    + \sum_i |\sigma_i| \left[ \varepsilon_0 + V(r_i) \right] , 
    \label{eq: toy model}
\end{eqnarray}
where $\Gamma_0$ is the quantum of circulation, $\sigma_i = 0,\pm 1$ corresponds to no-vortex, a vortex, or an antivortex in the $i$-th position of the lattice, $r_{ij}$ is the distance between the $i$-th and $j$-th vortices, $\varepsilon_0$ is the energy required to generate a vortex/antivortex in the bulk of the condensate, $V(r_i)$ is the trapping quadratic potential, and $N_c$ is the number of particles per cell.

The first term of the Hamiltonian corresponds to the interaction between vortices in 2D \cite{Onsager1949} (which are of long range in this problem \cite{Gifford2008}). The notation $\llangle ij \rrangle$ indicates that the sum is computed up to the fifth neighbours. This is done to avoid computing excessive long range interactions, but is also justified by the fact that we are interested in the role of defects in an already established lattice. The second term is the rotation energy, as $h N_c$ is approximately the angular momentum of a vortex. 
The parameter $\alpha$ accounts for the coupling of the vortices with the long range field generated by the remaining far away vortices. In the Bethe mean field approximation, the total order field is effectively $\Omega + \Omega'$, where $\Omega'$ is generated by those neglected long range interactions such that $\alpha \Omega \approx \Omega + \Omega' $. These coefficients are kept constant when varying $\Omega$, as the interaction of the BEC with the external field must be proportional to the angular momentum per vortex $J_z \sim M \Omega / N_v$ (where $M$ is the total condensate mass). Note that as the number of vortices $N_v$ grows linearly with the rotation speed, the coupling remains the same. Finally, the last term in Eq.~(\ref{eq: toy model}) corresponds to the energy required to pin a vortex in the system at a given point (including the vortex energy plus the trap potential energy). This Hamiltonian will be solved to obtain equilibria. Even though in the dynamical case (e.g., solving RGLE or SRGLE, or the Gross-Pitaevskii equation) vortices first appear near the border of the condensate, as time evolves they move inwards and in the equilibrium they remain at the center. We will thus only compare states generated by this model with steady state equilibria reached by SRGLE.

\begin{figure}
    \includegraphics[width=\columnwidth]{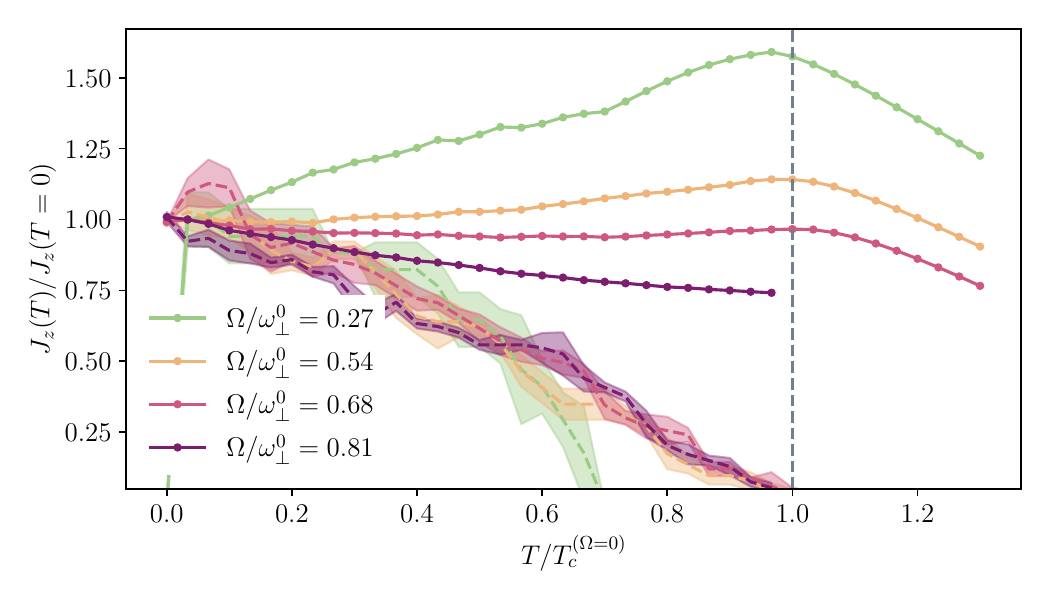}
    \caption{Solid lines: Angular momentum in the $z$ direction normalized by its value at $T=0$, as a function of $T$ for different values of $\Omega$ in RSGLE simulations. Dashed lines: Estimated angular momentum of quantized vortices, normalized by $J_z(T=0)$. The vertical dashed line indicates $T_c^{\Omega=0}$. The colored shaded area indicates $95\%$ confidence levels.}
    \label{Figure:angular momentum}
\end{figure}

To obtain equilibrium states of the Hamiltonian in the canonical ensemble at a given $T$ we use the Metropolis-Hastings algorithm \cite{Metropolis1953}. To reproduce the case of constant volume we vary the trapping potential with $\Omega$ in the same proportion as in the SRGLE simulations. The bottom row of Fig.~\ref{Figure:phase transition} shows the vortex lattice in the model for a given $\Omega$ and for increasing $T$, where white, blue, and red dots represent no-vortex, co-rotating, and counter-rotating vortices respectively. Despite the fact that the number of vortices in the model is larger than in the RSGLE simulations, the way the lattice melts and disorder increases is reminicent of the numerical simulations: the borders become increasingly disordered and the coherence length is lost from the borders to the center. 

The bottom panel in Fig.~\ref{Figure:vortex evolution} shows the net number of co-rotating vortices in Monte Carlo simulations of the vortex-energy model under ``constant volume.'' The model, that takes into account only vortex interactions and positional energies in the lattice, captures qualitatively features seen in the RSGLE simulations (Fig.~\ref{Figure:vortex evolution}, top panel). For increasing $\Omega$ indeed more vortices remain at a fixed $T$, even for temperatures close to $T^{\Omega=0}$, resulting in a positive shift in $T_c$ (note that, even though we are studying melting with the model, a shift to higher temperatures of the melting temperature implies a shift to higher temperatures of the Bose-Einstein critical temperature, and under our conditions both temperatures are similar \cite{Gifford2008}). Also, the overshooting in the number of vortices for low $T$ is captured by the model and, as in RSGLE runs, the effect is stronger for small $\Omega$. The latter effect arises here in the same way as in RSGLE: thermal fluctuations provide enough energy to facilitate the excitation of new vortices in the condensate, specially near the border and near the critical value of $\Omega$ to create the first vortex. This, as in RSGLE simulations, makes new vortex states in phase space available for the system to explore as $T$ increases. The overshooting suggests that the increase in the vortex number could be associated to a less energetic state for zero temperature which was separated from the original by some energy barrier that the system could not overcome. 

\section{Angular momentum}

At this point features of the array melting process seem to be captured by the vortex-energy model, but other aspects of the transition remain elusive. Moreover, the mere disappearance of vortices only partially addresses our questions. Are vortices truly disappearing, or are they engulfed by thermal noise? How does the system as a whole respond as $T$ increases, and what role does the non-condensed gas play? To investigate these questions we study the angular momentum $J_z$ in the RGLE simulations. This quantity depends on the spatial mass distribution in the condensed and thermalized phases, as well as on their respective velocity fields, allowing us to consider the whole system. Figure \ref{Figure:angular momentum} shows in solid lines the total $J_z$ as a function of $T$ for various $\Omega$ in the constant volume case, and in dashed lines the estimated contribution to $J_z$ from quantized vortices (i.e., only from the condensate), both normalized by $J_z$ at $T=0$. The angular momentum of quantized vortices is computed by multiplying the angular momentum per vortex in the fundamental state by the total number of vortices in each state. For rapid rotation and in the Thomas-Fermi approximation, the angular momentum per vortex is constant, independent of $\Omega$, and $\approx 2 N \hbar/7$ (where $N$ is the total number of particles in the condensate). The theoretical value and other estimations from the simulations are close to each other, so we consider this to be a good approximation of $J_z$ in the condensed phase.

At low $T$, all angular momentum is in the lattice. For low rotation speeds, total $J_z$ increases with $T$ due to the rising fraction of normal fluid. Fluctuations are more significant at the periphery of the condensate, where the normal fluid accumulates, leading also to an expansion of the system radius. Additionally, this region experiences greater inertial forces, causing the normal fluid to rotate, and resulting in the observed increase in total $J_z$ with $T$. Note that the thermalized gas can contribute angular momentum without the need for additional vortices in the lattice, and can even outweigh the loss of $J_z$ in the condensate due to their disappearance. This effect is evident in the slow rotating cases, such as $\Omega /\omega_{\perp}^0 = 0.27$ and $0.54$, where total $J_z$ grows almost linearly with $T$, while the amount of vortices and of $J_z$ in the condensate drop dramatically. As the critical temperature is approached, temperature fluctuations become so pronounced that they decorrelate any other effects, causing $J_z$ to drop rapidly. At larger $\Omega$ the behavior becomes less pronounced: at $\Omega /\omega_{\perp}^0 = 0.68$ total $J_z$ is almost constant for $T<T_c^{\Omega=0}$, meaning that the loss of $J_z$ from vortices is compensated by the normal fluid, while at $\Omega /\omega_{\perp}^0 = 0.81$ it only decreases. This can be explained by the ability of the condensate to move vortices from the core of the condensate towards the perifery for large $\Omega$, impairing the normal fluid of generating angular momentum. This is particularly noticeable up to $T/T_c^{\Omega=0} \approx 0.3$ where angular momentum due to the vortices and the total $J_z$ are close to each other. Afterwards, the number of vortices decreases and there is enough space for the normal fluid to generate $J_z$. Note also that as the vortex-energy model reproduces the behavior of the number of vortices, it also reproduces the general behavior of $J_z$ from the condensate seen in Fig.~\ref{Figure:angular momentum}. 

\section{Conclusions}

We showed that rotation induces a positive shift in the critical temperature of a BEC, provided that the volume of the condensate remains constant. When this is not the case (with fixed potential) the broadening of the condensate cloud leads to a decrease of mass density, outweighing the positive effect of rotation, and leading to a net negative shift of the critical temperature. Second, we introduced a vortex-energy model that accurately reproduced the critical temperature behavior in the fixed-volume case, indicating that this shift is driven by interactions between vortices and their positional energy. Thus, the rigidity of the vortex lattice provides a long-range order that allows the condensate to persist at higher temperatures.
We also studied the vortex lattice melting process, which occurs from the edge inward, accompanied by cloud broadening -- an effect consistent with our model. Finally, we showed that the relationship between angular momentum and temperature is highly dependent on the rotation rate, and related it with the vortex number and the appearance of thermalized fluid. 

The proposed model could be further utilized to understand the transition of the condensate to a normal fluid through a two-fold perspective: a typical BEC transition coupled with the two-dimensional melting of a quantum vortex lattice. While the former has been extensively studied, the latter has received less attention, and further investigation through simulations in elongated traps could provide deeper insights into this transition.

\begin{acknowledgments}
The authors acknowledge support from UBACyT Grant No.~20020220300122BA, and from proyecto REMATE of Redes Federales de Alto Impacto, Argentina. 
\end{acknowledgments}

\appendix

\begin{figure}
\centering
\includegraphics[width=\columnwidth]{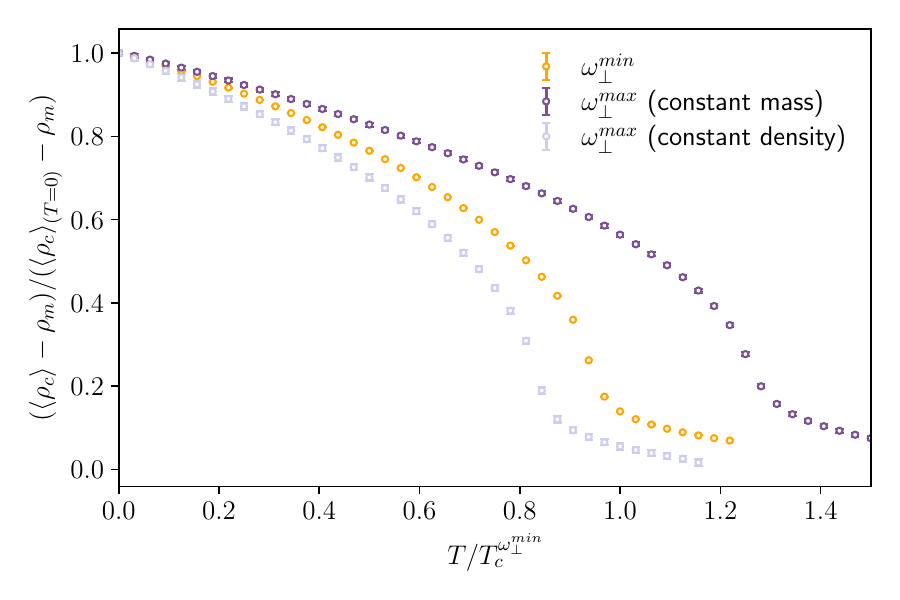}
\caption{Density at the center of the trap minus the mean density in the trap, normalized by their values at $T = 0$, and as a function of the temperature. Three cases are shown with $\Omega=0$: $\omega_{\perp}^{min}$ corresponds to the minimum trap frequency used in this study, $\omega_{\perp}^{max}$ is the maximum frequency used (with the same total mass in the trap as when using $\omega_{\perp}^{min}$), and $\omega_{\perp}^{max}$ (constant density) corresponds to a case in which the density was kept the same. Temperatures are normalized by the critical temperature for $\omega_{\perp}^{max}$.} 
\label{Figure: critical temperatures}
\end{figure}

\section*{Appendix: Effect of varying the external potential in $T_c$ \label{sec: appendix}}

For a non-rotating case, three temperature scans were performed using the SRGLE with $\Omega=0$: the first using the smaller trapping frequency considered in this paper, the second with the largest (both with the same total mass), and a third with the largest frequency but modifying the total mass so that the mean density in the condensate remained the same (within $1.3 \%$ accuracy). The mean density was computed using the radius of the condensate that follows from the Thomas-Fermi approximation, and also taking the distance from the origin at which the mass dropped below a fixed threshold, and in both cases we obtained similar estimations for the density. The results of the three temperature scans are shown in Fig.~\ref{Figure: critical temperatures}. Increasing the trap frequency without any other constraint results in an increase of the critical temperature, but increasing the trap frequency while maintaining the same density results in a small decrease of the critical temperature. The effect is the opposite of the change in $T_c$ observed when changing rotation while increasing the trap frequency at a constant density.

\bibliography{ms}

\end{document}